\documentstyle[12pt]{article}

         \def\ba{\begin{array}}
         \def\ea{\end{array}}
         \def\be{\begin{equation}}
         \def\bea{\begin{eqnarray}}
         \def\eea{\end{eqnarray}}
         \def\ee{\end{equation}}

         \def\bq{\begin{eqnarray}}
         \def\eq{\end{eqnarray}}
         
         \def\O{{\cal O}}
\begin{document}
\hfill
\vbox{
    \halign{#\hfil         \cr
            hep-th/9611234      \cr
           } 
      }  
\vskip 10 mm
\centerline{ \Large \bf
        Derivation of quantum theories:  }
\centerline{ \Large \bf
        symmetries and the exact solution of the derived system}
\vskip 25 mm
\centerline{\bf M. Khorrami$^{a,b,c}$\footnote {e-mail:mamwad@netware2.ipm.ac.ir},
A. Aghamohammadi$^{b,d}$\footnote {e-mail:mohamadi@netware2.ipm.ac.ir},
   M. Alimohammadi$^{a,b}$\footnote {e-mail:alimohmd@netware2.ipm.ac.ir}}
\vskip 20 mm
{\it
\centerline{ $^a$ Department of Physics, Tehran University,
             North-Kargar Ave.}
\centerline{ Tehran, Iran. }
\centerline{ $^b$ Institute for Studies in Theoretical Physics and
             Mathematics,}
\centerline{ P.O.Box  5531, Tehran 19395, Iran. }
\centerline{ $^c$ Institute for Advanced Studies in Basic Sciences,
             P.O.Box 159,}
\centerline{Gava Zang, Zanjan 45195, Iran. }
\centerline{ $^d$ Department of Physics, Alzahra University,
             Tehran 19834, Iran }
  }
\vskip 4cm
\begin{abstract}
Based on the technique of derivation of a theory, presented in our recent
paper \cite{KA}, we investigate the properties of the derived quantum
system. We show that the derived quantum system possesses the (nonanomalous)
symmetries of the original one, and prove that the exact Green functions of
the derived theory are expressed in terms of the semiclassically
approximated Green functions of the original theory.
\end{abstract}
\newpage

\section{\bf Introduction}
In recent decades some two-dimensional exactly solvable quantum field
theories have been discovered. 2-dimentinall conformal field theory
\cite{BPZ} and 2-dimensional Yang-Mills theory \cite{W} are two important
examples of them. There are also some recent advances \linebreak in $d>2$
dimensions, where the 4-dimensional supersymmetric gauge theories is one of
the most important ones \cite{SW}.

In this paper we want to discuss a class of exactly solvable quantum
theories. In the paper \cite{KA}, it was shown that, starting from an
action, one can construct other actions which the classical behaviour of
them are deduced from those of the original action. The procedure relating
the original action to the new one is just a derivation. Now we want to
continue our investigation by studying the behaviour of a quantum
system obtained by derivation from an action of a quantum theory.
It turns out that the derived system possesses all of the (nonanomalous)
symmetries of the original system, just as the case of classical systems.
Here, however, a novel property arises: the derived quantum theory is
{\it almost classical}; that is in the derived theory there are only
one loop quantum corrections to the classical action. Using this property,
one can calculate all of the Green functions of the derived theory exactly,
even though this may be not the case for the original theory. Note that
the classical equations of motion are the same in the original theory and
the derived one, although their quantum behaviour are completely different.

The paper consists of three sections. In section 1, we show that the derived
quantum system possesses the (nonanomalous) symmetries of the original
theory. In section 2, we study a specific example: a Coulomb gas theory,
which has conformal symmetry. We show that the derived theory is a
combination of two theories : a Coulomb gas theory which its central charge is
related to the derivative of that of the first and a free bosonic theory. Finally, in
section 3 we prove that the generating functional of the connected diagrams
of  the derived theory has only tree- and one loop-contributions, and these
have simple relations to the tree- and one loop-parts of the generating
functional of the connected diagrams of the original theory. So, solving the
original theory up to one loop allows one to solve the derived theory
exactly.

\section{\bf Action formulation, preservation of symmetries}
Consider the action
\be\label{99} S^{(0)}=S(\phi ,\lambda ),\ee
where $\phi$ denotes the dynamical variable(s) of the system and $\lambda$
is some parameter. As in \cite{KA}, differentiating this action with respect
to $\lambda$, treating $\phi$ as a function of $\lambda$, leads to:
\be\label{98} S^{(1)}:={{dS^{(0)}}\over{d\lambda}}=
    {{\partial S}\over{\partial\lambda}}+\Big<{{\delta S}\over{\delta\phi}},
    \psi\Big> ,\ee
where we have defined
\be \psi :={{d\phi}\over{d\lambda}}.\ee
In \cite{KA}, it was shown that if the first action has a symmetry, the
derived action $S^{(1)}$, possesses this symmetry as well. We now want to
show that if the symmetry of the first theory is not anomalous, the
symmetry of the derived one is not anomalous either.

To do this, one must show that the path integral measure corresponding to
the derived theory does not change under the symmetry transformation. We
know that, under the symmetry transformation
\be \phi_\lambda\to\O_\lambda\phi_\lambda ,\ee
the measure does not change:
\be D\phi_\lambda =D(\O_\lambda\phi_\lambda ),\ee
since the symmetry of the first theory has no anomaly. This shows that
the symmetry of the theory defined through the action
\be\label{89} S^{(1)\Delta}:={1\over\Delta}\Big[ S(\phi_{\lambda +\Delta},
\lambda +\Delta )-S(\phi_\lambda ,\lambda )\big] ,\ee
is nonanomalous. Now consider the following change of variables
\bea
   \psi^\Delta&:=&{1\over\Delta}(\phi_{\lambda +\Delta}-\phi_\lambda )
   \cr\label{88}\phi&:=&\phi_\lambda .\eea
The Jacobian of this transformation is a constant:
\be D\psi^\Delta\; D\phi =\Big({1\over\Delta}\Big)^N D\phi_{\lambda +\Delta}
    \; D\phi_\lambda .\ee
This shows that, under the transformation
\bea \label{1}\phi&\to&\O_\lambda\phi\cr
       \psi^\Delta&\to&{1\over\Delta}[\O_{\lambda +\Delta}
       (\phi +\Delta\psi^\Delta )-\O_\lambda\phi ] ,\eea
the measure corresponding to $\phi$ and $\psi^\Delta$ does not change:
\be D\psi^\Delta\; D\phi =D\Big\{ {1\over\Delta}[\O_{\lambda +\Delta}
    (\phi +\Delta\psi^\Delta )-\O_\lambda\phi ]\Big\} D(\O_\lambda\phi ).\ee
Now let $\Delta$ tends to zero. The transformation (\ref{1}) becomes
\bea \phi&\to&\O_\lambda\phi =:\O\phi\cr
\psi&\to&{d\over{d\lambda}}(\O_\lambda\phi )=\Big({{\partial\O_\lambda}
\over{\partial\lambda}}\phi +{\partial\over{\partial\phi}}(\O_\lambda\phi )
\psi\Big) =:\O\psi ,\eea
and therefore
\be D\phi\; D\psi =D(\O\phi )\; D(\O\psi ).\ee
This completes the proof.

\section{\bf A simple example: derivation of a Coulomb gas theory} As a
simple example, consider a Coulomb gas theory with the following
energy--momentum tensor \cite{DF}
\be\label{2} T^{(0)}(z)= -{1\over 2}:\partial\phi\partial\phi +{Q\over 2}
\partial^2\phi .\ee
It is well known that this is a conformal field theory (CFT) with the
central charge
\be c^{(0)}=1+3Q^2,\ee
and the following OPE of the fields
\be \phi (z)\phi (w)=-\ln (z-w)+\hbox{ regular terms}.\ee
Differentiating (\ref{2}) with respect to $Q$, we obtain
\be T^{(1)}(z)= -:\partial\phi\partial\psi :+{Q\over 2}
\partial^2\psi +{1\over 2}\partial^2\phi .\ee
As this component of the energy-momentum tensor is independent of $\bar z$,
the derived theory is conformal as well. Now, if $T^{(1)}$ is to be the
energy--momentum tensor of a new theory, which contains the fields $\phi$
and $\psi$, we must have the following OPE's
\bea T^{(1)}(z)\phi (w)&=&{{\partial\phi (w)}\over{z-w}}+
\hbox{ terms containing no derivative} + \hbox{ regular terms}\cr
       T^{(1)}(z)\psi (w)&=&{{\partial\psi (w)}\over{z-w}}+
\hbox{ terms containing no derivative} + \hbox{ regular terms}.\eea
This relations are satisfied, provided $\phi (z)\phi (w)$ and
$\psi (z)\psi (w)$ are regular and
\be\phi (z)\psi (w)=-\ln (z-w)+ \hbox{ regular terms},\ee
that is, provided we have
\be\label{5} <\phi (z)\phi (w)>=<\psi (z)\psi (w)>=0,\ee
and
\be\label{6} <\phi (z)\psi (w)>=-\ln (z-w).\ee
(\ref{5}) and (\ref{6}) are a special case of the correlation functions of
a derived theory to be discussed in the next section.

Using the above-mentioned OPE's, it is readily seen that
\be T^{(1)}(z)T^{(1)}(w)={{1+3Q}\over{(z-w)^4}}+
   {{2 T^{(1)}(w)}\over{(z-w)^2}}+{{\partial T^{(1)}(w)}\over{z-w}}+
   \hbox{ regular terms},\ee
which shows that $T^{(1)}$ is indeed the energy-momentum tensor of a
CFT with the central charge
\be c^{(1)}=2+6Q=2+{{dc^{(0)}}\over{dQ}}.\ee
It can be also shown that the fields
\bea V_1:=\phi -Q\psi\cr V_2:=\partial V_1\eea
are the primary fields of this theory with the conformal weights
$h_1=0$ and $h_2=1$, respectively. Also, one can write $T^{(1)}$ in terms of
two new fields
\bea \chi_1:={1\over{\sqrt{2Q}}}(\phi -Q\psi)\cr
     \chi_2:={1\over{\sqrt{2Q}}}(\phi +Q\psi)\eea
as
\be T^{(1)}(z)={1\over 2}:\partial\chi_2\partial\chi_2
  -{1\over 2}:\partial\chi_2\partial\chi_2 +{Q'\over 2}\partial^2\chi_2 .\ee
where
\be Q':=\sqrt{2Q}.\ee
The fields $\chi_1$ and $\chi_2$ satisfy the following OPE's:
\bea \chi_1(z)\chi_1(w)=\ln (z-w)+ \hbox{ regular terms},\cr
     \chi_2(z)\chi_2(w)=-\ln (z-w)+ \hbox{ regular terms},\cr
     \chi_1(z)\chi_2(w)= \hbox{ regular terms}.\eea
So $\chi_1$ and $\chi_2$ are two independent fields, and the central charge
of the derived theory is the sum of two central charges: the central charge
of a free bosonic theory ($c=1$), and that of a Coulomb gas theory
($c=1+3Q'^2=1+6Q$), which is the same as the result obtained earlier.
One can continue the derivation. After $n$ times derivation, we will have
\be T^{(n)}={{d^nT}\over{dQ^n}},\ee
which is the energy momentum tensor of a CFT with central charge
\be c^{(n)}=n+1+{{d^nc^{(0)}}\over{dQ^n}}.\ee

\section{\bf Exact form of the generating functional of a derived action}
Consider an action $S^{(0)}(\phi )$ of the form (\ref{99}), which contains
no first order terms in $\phi$, and its derived action
$S^{(1)}(\phi ,\psi )$, which is of the form (\ref{98}). The generating
functional corresponding to such an action is
\be Z^{(1)}(J, j):=\int D\phi\; D\psi\;\exp\left[{i\over\hbar}\left(
<{{\delta S^{(0)}}\over{\delta\phi}}, \psi > +{{\partial S^{(0)}}
\over{\partial\lambda}}-<j, \psi > - <J, \phi >\right)\right] .\ee
Integrating over $\psi$, and then $\phi$, we obtain
\bea Z^{(1)}(J, j)&:=&\int D\phi\;\delta\left(
{{\delta S^{(0)}}\over{\delta\phi}}-j\right)\exp\left[{i\over\hbar}\left(
{{\partial S^{(0)}}\over{\partial\lambda}} - <J, \phi >\right)\right] \cr
&=&\left\{\left[ {\rm det}\left(
{{\delta^2 S^{(0)}}\over{\delta\phi^2}}\right)\right]^{-1}
\exp\left[{i\over\hbar}\left({{\partial S^{(0)}}\over{\partial\lambda}}
- <J, \phi >\right)\right]\right\}\vert_{\phi =\phi (j)},\eea
where $\phi (j)$ is the solution of the classical equation of motion of
$\phi$ with the source $j$:
\be {{\delta S^{(0)}}\over{\delta\phi}}\vert_{\phi (j)} =j.\ee
From $Z^{(1)}$, one obtains the generating functional of connected diagrams:
\be W^{(1)}:=\ln Z^{(1)}\qquad\Rightarrow\nonumber\ee
\bea W^{(1)}(J, j)&=& -{i\over\hbar}<J, \phi (j)> +
{i\over\hbar}{{\partial S^{(0)}(\phi (j))}\over{\partial\lambda}}
- {\rm tr}\;\ln\left[{{\delta^2 S^{(0)}}\over{\delta\phi^2(j)}}\right]
\cr\label{97} &=:&\label{95} W^{(1)}_{(0)}(J, j)+
W^{(1)}_{(1)}(J, j),\eea
where the subscripts refer to the number of loops. So the exact generating
functional has only zero- and one-loop contributions. We can compare this to
the generating functional of the original theory, up to the first loop
order. We have
\be Z^{(0)}(j):=\int D\phi\;\exp\left[{i\over\hbar}\left(
S^{(0)} - <j, \phi >\right)\right] .\ee
Expanding $\phi$ around $\phi (j)$, and keeping only terms up to second
order, we obtain
\bea Z^{(0)}(j)&:=&\int D\phi\;\exp \left\{ {i\over\hbar}\left[ S^{(0)}[
\phi (j)]+{1\over 2}<\phi -\phi (j), {{\delta^2 S^{(0)}}\over{\delta\phi^2}}
\vert_{\phi (j)}[\phi -\phi (j)]> - <j, \phi (j)>\right]\right\}\cr
&=&\exp \left\{ {i\over\hbar}\left[ S^{(0)}[\phi (j)] - <j, \phi (j)>
\right]\right\}\left[{\rm det}\left({{\delta^2 S^{(0)}}\over
{\delta\phi^2(j)}}\right)\right]^{-1/2},\eea
and from this,
\bea W^{(0)}_1(j)&=& {i\over\hbar}\left\{ S^{(0)}[\phi (j)] -<j, \phi (j)>
\right\} -{1\over 2}{\rm tr}\;\ln
\left[{{\delta^2 S^{(0)}}\over{\delta\phi^2(j)}}\right]\cr
\label{96} &=:& W^{(0)}_{(0)}(j)+W^{(0)}_{(1)}(j).\eea
Comparing this with (\ref{97}), we see that
\be\label{94} W^{(1)}_{(0)}(J, j)=<J,{{\delta W^{(0)}_{(0)}(j)}\over
{\delta j}}> + {{d W^{(0)}_{(0)}(j)}\over{d\lambda}},\ee
and
\be W^{(1)}_{(1)}(J, j)=W^{(1)}_{(1)}(j)=2W^{(0)}_{(1)}(j).\ee
Note that (\ref{96}) is an approximation of the generating functional of
the original theory, whereas (\ref{95}) is the exact generating functional
of the derived one.

From (\ref{94}), it is easily seen that
\be {{\delta^2 W^{(1)}}\over{\delta J^2}}=0,\ee
which means that
\be <\phi_1\phi_2 \cdots >^{(1)}_c=0,\ee
where the dots denote any combination of $\phi$'s and $\psi$'s, and the
superscript refers to the derived theory. We also have
\be i\hbar{{\delta W^{(1)}}\over{\delta J}}=
    i\hbar{{\delta W^{(0)}_{(0)}}\over{\delta j}}.\ee
So
\be (i\hbar)^{n+1}{{\delta^{n+1} W^{(1)}}\over{\delta j^n \delta J}}=
    (i\hbar)^{n+1}{{\delta^{n+1} W^{(0)}_{(0)}}\over{\delta j^{n+1}}},\ee
which means
\be\label{85}<\phi_0\psi_1\cdots\psi_n>_c^{(1)}=
<\phi_0\cdots\phi_n>_{c,(0)}^{(0)};\ee
that is, the exact connected Green function of one $\phi$ and $n$ $\psi$'s
is equal to the Green function of $n+1$ $\phi$'s in the original theory at
the tree level.

Finally,
\be (i\hbar)^n{{\delta^n W^{(1)}}\over{\delta j^n}}\Big\vert_{J=0}=
    2(i\hbar)^n{{\delta^n W^{(0)}_{(1)}}\over{\delta j^n}}+
 (i\hbar)^n{d\over{d\lambda}}{{\delta^n W^{(0)}_{(0)}}\over{\delta j^n}},\ee
which means
\be\label{84}<\psi_1\cdots\psi_n>_c^{(1)}=
2<\phi_1\cdots\phi_n>_{c,(1)}^{(0)} +
{d\over{d\lambda}}<\phi_0\cdots\phi_n>_{c,(0)}^{(0)}.\ee

To summerize, we showed that one can calculate all of the exact connected
Green functions of the derived theory in terms of the connected Green
functions of the original theory up to one loop.

There exists another proof for this result. Consider the transformation
(\ref{88}) and the definition (\ref{89}). It is seen that
\bea <\phi_0\cdots\phi_k\psi_{k+1}^\Delta\cdots\psi_n^\Delta>^{(1)\Delta}_c
&=&{{(-1)^{n-k}}\over{\Delta^{n-k}}}<\phi_0\cdots\phi_n>_c^{(1)\Delta}
\cr\label{87} &=&{{(-1)^{n-k}}\over{\Delta^{n-k}}}
<\phi_0\cdots\phi_n>_{c,-\Delta\hbar}^{(0)},\eea
where the last relation means the $n+1$ point function calculated using
$-\Delta\hbar$ instead of $\hbar$.
Now we have
\be\label{86} <\phi_0\cdots\phi_n>_{c,-\Delta\hbar}^{(0)}=
    <\phi_0\cdots\phi_n>_{c,(0),-\Delta\hbar}^{(0)}+
    <\phi_0\cdots\phi_n>_{c,(1),-\Delta\hbar}^{(0)}+\cdots ,\ee
which is an expansion in terms of powers of $-\Delta\hbar$. The zeroth term
is of the order $(-\Delta\hbar)^n$. It is seen that if $k>0$, the limit of
the right-hand side of (\ref{87}) at $\Delta\to 0$ is zero, and if $k=0$,
only the zeroth term of the right-hand side of (\ref{86}) survives in the
limit, which yields (\ref{85}).

To calculate the $n$ point function of $\psi$'s, we note that
$$ <\psi_1\cdots\psi_n>_c^{(1)\Delta}={1\over{\Delta^n}}\left[
     <\phi_1\cdots\phi_n>_{c,\Delta\hbar}^{(0)}\vert_{\lambda +\Delta} +
     (-1)^n<\phi_1\cdots\phi_n>_{c,-\Delta\hbar}^{(0)}\vert_\lambda\right]$$
$$ ={1\over{\Delta^n}}\left[
   <\phi_1\cdots\phi_n>_{c,(0),\Delta\hbar}^{(0)}\vert_{\lambda +\Delta} +
   <\phi_1\cdots\phi_n>_{c,(1),\Delta\hbar}^{(0)}\vert_{\lambda +\Delta} +
   \cdots\right. $$
\be\left. +(-1)^n<\phi_1\cdots\phi_n>_{c,(0),-\Delta\hbar}^{(0)}
   \vert_\lambda +(-1)^n<\phi_1\cdots\phi_n>_{c,(1),-\Delta\hbar}^{(0)}
   \vert_\lambda +\cdots\right] .\ee
The first term in $<\phi_1\cdots\phi_n>_{c,\Delta\hbar}^{(0)}$ is of order
$(\Delta\hbar )^{n-1}$. So
\be <\psi_1\cdots\psi_n>_c^{(1)}=\lim_{\Delta\to 0}
{{<\phi_1\cdots\phi_n>_{c,(0)}^{(0)}\vert_{\lambda +\Delta}-
<\phi_1\cdots\phi_n>_{c,(0)}^{(0)}\vert_\lambda}\over\Delta}+
2<\phi_1\cdots\phi_n>_{c,(1)}^{(0)},\ee
which is precisely (\ref{84}).

One can contninue the procedure and differentiate further. In a similar
manner, it turns out that only zero- and one-loop terms contribute to
the Green functions.
\vskip 1cm
\noindent{\bf Acknowledgement} M. Khorrami and M. Alimohammadi would like to
thank the research vice-chancellor of the university of Tehran, the work
was partially supported by them.
\vskip 2cm

\end{document}